\documentclass[a4paper,11pt]{article}
\usepackage[affil-it]{authblk}
\usepackage[adobe-utopia]{mathdesign}
\usepackage{amsmath,amsthm,multirow}
\usepackage{color}
\newtheorem{observation}{Observation}
\newtheorem*{observation*}{Observation}
\usepackage{cite}
\usepackage{hyperref}

\usepackage[explicit]{titlesec}
\renewcommand\thesection{\Roman{section}}
\renewcommand\thesubsection{\Alph{subsection}}
\renewcommand\thesubsubsection{\arabic{subsubsection}}
\titleformat{\section}
{\centering\normalfont\normalsize}{\thesection. \MakeUppercase{#1}}{1em}{}
\titleformat{\subsection}
  {\centering\normalfont\normalsize}{\thesubsection. #1}{1em}{}
\titleformat{\subsubsection}
  {\centering\normalfont\normalsize\itshape}{\thesubsubsection. #1}{1em}{}

\def\princ{\mathbf{A}}
\def\evac{\mathbf{E}}
\def\wtildep{\tilde w^\princ}
\def\wtildee{\tilde w^\evac}
\def\ctilde{\tilde c}

\begin{document}
\title{The Cry Wolf Effect in Evacuation: a Game-Theoretic Approach\thanks{Accepted for publication in Physica A. DOI: \href{https://doi.org/10.1016/j.physa.2019.04.126}{10.1016/j.physa.2019.04.126}}}

\author[1]{Alexandros Rigos}
\author[1]{Erik Mohlin}
\author[2\thanks{Corresponding author. Email: enrico.ronchi@brand.lth.se.}]{Enrico Ronchi}
\affil[1]{Department of Economics, Lund University, P.O. Box 7082, SE-220 07, Lund, Sweden}
\affil[2]{Department of Fire Safety Engineering, Lund University, P.O. Box 118, SE-221 00 Lund, Sweden}
\date{}%
\maketitle%
\begin{abstract}%
In today's terrorism-prone and security-focused world, evacuation emergencies, drills, and false alarms are becoming more and more common. Compliance to an evacuation order made by an authority in case of emergency can play a key role in the outcome of an emergency. In case an evacuee experiences repeated emergency scenarios which may be a false alarm (e.g., an evacuation drill, a false bomb threat, etc.) or an actual threat, the Aesop's cry wolf effect (repeated false alarms decrease order compliance) can severely affect his/her likelihood to evacuate. To analyse this key unsolved issue of evacuation research, a game-theoretic approach is proposed. Game theory is used to explore mutual best responses of an evacuee and an authority. In the proposed model the authority obtains a signal of whether there is a threat or not and decides whether to order an evacuation or not. The evacuee, after receiving an evacuation order, subsequently decides whether to stay or leave based on posterior beliefs that have been updated in response to the authority's action. Best-responses are derived and Sequential equilibrium and Perfect Bayesian Equilibrium are used as solution concepts (refining equilibria with the intuitive criterion). Model results highlight the benefits of announced evacuation drills and suggest that improving the accuracy of threat detection can prevent large inefficiencies associated with the cry wolf effect.
\end{abstract}%

\section{Introduction}

After every evacuation associated with a terrorist attack, a fire, a natural disaster or a human error \cite{Averill}, investigators struggle to find out if decisions made by authorities or evacuees was appropriate. In some cases, experts find that a quicker or different response to a threat could have saved a tremendous number of lives. At the same time, the authorities ordering an evacuation and the evacuees have the difficult task of taking decisions with time pressure and often with only scarce information available. 
To date, many tools and models have been developed to represent decision making during evacuation in the context of applied physics research. They include, for example, simulators of pedestrian dynamics during evacuation \cite{Helbing2000,Seyfried,Lovreglio15}, which aims at representing human movement and associated behaviour in case of different scenarios and threats \cite{Isobe,Cirillo}, or models of authority's recommendations \cite{Kyne}. Human behaviour in evacuation scenarios has also been investigated using Virtual Reality tools, as they allow the investigation of individual and group decision making during evacuation \cite{Moussaid2016,Kinetader,Ronchi2016a,Bode}. Experimental work has been performed in order to validate such models and tools, including the study of different types of emergent behaviours related to evacuation dynamics \cite{Hoogendoorn,Seyfried2007,Ronchi2016b,Lovreglio2014,Bellomo}.

The main limitations of these tools is that they consider evacuation scenarios in isolation and they address only one of the two parties involved in an evacuation, either (i) the decision making of the authority (i.e. their orders/instructions) or (ii) the actions of the evacuees. In contrast, currently there is no framework available which is able to comprehensively consider optimal decision making strategies for an authority and an evacuee at the same time in case of several repeated emergency evacuation threats. 

Many tragedies have demonstrated the dilemmas that decision makers may face when dealing with emergency situations. For instance, in relation to the Costa Concordia disaster, the Italian court trial is evaluating the behaviour of captain Schettino who allegedly did not order the evacuation of the ship on time \cite{Kvamme}. Apart from the Captain's negligence or a simply incorrect assessment of the situation, the analysis of the trade-offs between the cost of a useless evacuation (i.e., the risk of ``ruining'' the holidays to the cruise ship guests for no reason due to a false alarm in the Costa Concordia case) and the cost of a delayed evacuation have not been accounted for appropriately. Similar challenges are present in case of emergency evacuations in other sensitive facilities. A known example is the case of the 2001 World Trade Center terrorist attacks in which instructions to remain in the stricken building were given to employees working in the WTC during the attack \cite{Kuligowski11}. In contrast, survivors' accounts demonstrated that in some instances an independent evacuee's decision made in conflict with the order received, can save lives.

On the one hand, authorities should not ignore bomb threats given the risk of losing human lives. On the other hand, multiple unnecessary evacuations (drills, false threats) may lead to a decrease in the authority credibility and people's underestimation of threats \cite{Gwynne16,Gwynne17}. This is called the cry wolf effect in reference to Aesop's fable \emph{The Boy who Cried Wolf} in which instruction compliance is decreased by a series of false alarms \cite{Breznitz}. Repeated unnecessary evacuations also have a high cost to society, including the risk of letting terrorists achieve their goal, \emph{i.e.} making people live in fear. Ethical dilemmas associated with the assessment of the cost-effectiveness of decision making in scenarios which involve human lives at risk have been discussed in different research fields such as road safety \cite{deBlaeij} or earthquake engineering \cite{Wu}. Although the cost of a life has been discussed in different contexts \cite{Viscusi,Hultkrantz}, an evaluation of decision making from both the point of view of the evacuee and the authority has not been conducted systematically in case of emergency evacuation. This is a rising issue, since in a terrorism-prone world, false bomb threats with subsequent evacuations are becoming more and more common \cite{Ronchi}. 

The concept of crowd panic in case of evacuation is often employed in the media. In the evacuation research community, it is instead known that panic rarely occurs in crowd evacuations \cite{Cocking}. Following the debate of the panic misconception, emergency evacuation research has focused on the application of different theories and models to increase the understanding of isolated decisions. This includes rationality-based theories of choice such as utility theories \cite{Lovreglio15}, the theory of affordances \cite{Gibson} psychological models \cite{Canter,Sime}, data-driven \cite{Johansson} or cognitive science-inspired models \cite{Moussaid11}. These models have shown reasonably good capabilities in capturing human behaviour and decision making during evacuations. Nevertheless, their major limitations are the boundaries of the system that they consider, \emph{i.e.} a single evacuation, a single group of evacuation scenarios in similar conditions or a specific behaviour in a given condition.

This paper proposes the use of a game-theoretic approach to systematically address evacuation emergencies. Game theory is suggested here given its unique ability to interpret the interactions between different intelligent decision makers and the opportunity to investigate mutual best responses in complex systems \cite{Marsan}.

Previous attempts to employ game theory to investigate decision making in emergencies have shown its great potential, but they mostly focused on isolated decisions of only one of the parties involved. For instance, interesting works on game-theoretic applications have been developed to study decisions taken by evacuees, \emph{e.g.} exit selection \cite{Ehtamo}, pedestrian movement and route choice \cite{Bellomo,Bouzat,Eng,Heliovaara,Shi}, or helping behaviors \cite{Moussaid16}. Similarly, single authority decisions in case of evacuation have been investigated using game theory, \emph{e.g.,} security prevention \cite{Reniers}, or risk perception \cite{Haynes}. Nevertheless, none of the previous studies have investigated systematically the interactions between the decisions of the authorities and the actions taken by the evacuees. The authority is here intended as the decision maker \emph{(e.g.} event safety manager, police officers, \emph{etc.)} who should decide whether or not to order an evacuation in case of a perceived threat. 

In this paper, we investigate the decision making process in repeated emergency evacuation with a static game-theoretic approach (Evacuation Game) aiming at studying the impact of different strategies adopted by an authority ($\princ$) and a potential evacuee ($\evac$) in a situation which involves uncertainty about the state of nature \emph{(i.e.} whether there is an actual threat or not). The evacuee has the option to stay or leave and the authority can decide whether or not to order an evacuation. The payoffs reflect the costs from the perspective of both $\princ$ and $\evac$.

\section{The Evacuation Game model}

Two players are considered in the Evacuation Game: An authority ($\princ$) and an evacuee ($\evac$). There are two states of nature, namely the normal state $n$ and the threat state $t$. The normal state $n$ corresponds to a situation of no actual (man-made or natural) disaster being imminent. The threat state $t$ refers to the case in which an actual disaster is taking place. 

\subsection{States of Nature}

The players share a common prior belief $\pi\in(0,1)$ that the state of nature is $t$. Correspondingly, the common prior belief that the state of nature is $n$ is the probability $1-\pi$.

\subsection{Actions}

Each of the two players has two actions available:
	$\princ$ can choose to either Order an evacuation ($O$) or to Not order an evacuation ($N$), 
	$\evac$ can choose to either Leave ($L$) or Stay ($S$). However, $\evac$ is given this choice only after $\princ$ chooses $O$. If $\princ$ chooses $N$, then the only choice available to $\evac$ is to Stay ($S$). This captures a situation in which evacuees are not directly exposed to threat \emph{(e.g.} they are not in the proximity of a fire, so they would not take the decision to leave independently) and their decision making relies on their compliance (or non-compliance) with the instruction given by $\princ$.

	\subsection{Available information}

	The authority $\princ$ receives information about the state of nature. In particular, $\princ$ obtains information in the form of a signal $X$ with two possible realizations corresponding to the two possible states of nature, i.e. signal $x_n$ for state $n$ and signal $x_t$ for state $t$. Such signals represent information received from a technological system to detect the threat \emph{(e.g.} a smoke detector, a clue of an imminent threat from a camera video). The quality and reliability of the signal (called here signal precision) is expressed with a variable $\tau\in[1/2,1]$ ($\tau$ is not lower than 1/2 as this case would correspond to a relabeling of the signals). This means that if the state of nature is $n$, then $\princ$ receives the signal $x_n$ with a probability $\tau$ and the signal $x_t$ with a probability $1-\tau$. Similarly, when the state of nature is $t$, then $\princ$ receives a signal $x_t$ with a probability $\tau$ and $x_n$ with a probability $1-\tau$. The precision $\tau$ is assumed to be commonly known and it depends on the source(s) of information about the threat. We use $\gamma_t$ and $\gamma_n$ to denote the authority's posterior belief that the state is $t$ after receiving the signal $x_t$ and $x_n$, respectively (see equations \eqref{gamma_t} and \eqref{gamma_n}) according to Bayes' rule. The posterior belief $\gamma_t$ is the probability that the authority assigns to the actual state being a threat, after having received the signal $x_t$ from a technological system. The definition is analogous for $\gamma_n$ and $x_n$.

\begin{gather}
\gamma _{t}=\Pr (\omega =t|X=x_{t})=\frac{\pi \tau }{\pi \tau +(1-\pi)(1-\tau )}  \label{gamma_t} \\
\gamma _{n}=\Pr (\omega =t|X=x_{n})=\frac{\pi (1-\tau )}{\pi (1-\tau)+(1-\pi )\tau }  \label{gamma_n}
\end{gather}

Note that the evacuee can only observe the action of $\princ$ (which can also be a lack of action in the case of non-order to evacuate $N$). The Evacuee has no information concerning the signal of $\princ$ or the state of nature but is aware of the precision of the threat detection (to a certain extent). This assumption is made since the evacuee does not have the information available to the authority \emph{(e.g.} $\princ$ might have received information concerning a bomb threat, \emph{etc.} which is not known to the evacuee) and is assumed not directly exposed to the threat \emph{(e.g.} not in the proximity of a fire, \emph{etc.).} This latter assumption is made since it considers the most challenging scenario from an evacuation perspective.

\subsection{Timing}

Different structures of timing can be used to represent the interactions among $\evac$ and $\princ$. In this instance, we present the most common scenario, \emph{i.e.,} a sequential timing in which the action of the authority $\princ$ is evaluated by the evacuee $\evac$ to take his/her action. This may represent the case of the sounding of an alarm in a building or an order to evacuate given by staff (or the absence of those). The timing can therefore be represented as follows:

\begin{enumerate}
  \item The state of nature $\omega\in\{n,t\}$ is realised.
  \item Given $\omega$, the signal $X\in\{x_n,x_t\}$ is realised.
  \item Given the received signal, the Authority takes an action $s\in\{O,N\}$ to either order an evacuation $O$ or to not order an evacuation $N$.
  \item The Evacuee observes the action (or inaction) of the Authority. If $s=O$, the Evacuee forms a (posterior) belief $\alpha=\Pr(\omega=t|s=O)$ that the state of nature is $t$ using Bayes' rule.
  \item If $s=O$, the Evacuee takes a decision $a\in\{L,S\}$.
  \item If $s=N$, the Evacuee has no other option than to decide to stay $S$.
  \item Payoffs are realised (see section \ref{payoffs}).
\end{enumerate}

\subsection{Strategies}

A strategy $m$ of the Authority $\princ$ consists of two probabilities $(m_n,m_t)\in[0,1]^2$. The number $m_i$ is the probability with which the Authority $\princ$ orders an evacuation ($s=O$) after observing signal $x_i\in\{x_n,x_t\}$. The total probability with which the Authority gives the evacuation order $s=O$ is presented in equation \eqref{totalprob}.
\begin{equation}\label{totalprob}
M=m_n ((1-\pi)\tau+\pi(1-\tau))+m_t ((1-\pi)(1-\tau)+\pi\tau)
\end{equation}

Let the probability that the Authority observes $x_t$ be denoted by $y_t=(1-\pi)(1-\tau)+\pi\tau$ and let the probability that the Authority observes $x_n$ be denoted by $y_n=(1-\pi)\tau+\pi(1-\tau)$. Equation \eqref{totalprob} then becomes:

\begin{equation}\label{reform}
M=m_n y_n+m_t y_t
\end{equation}

The strategy of the Evacuee is a number $r\in[0,1]$ expressing the probability with which the Evacuee chooses $L$ after the Authority has ordered an evacuation $O$. 

\subsection{Payoffs}\label{payoffs}

The payoff matrices presented in Table \ref{tab:payoffs}
describe the costs to the Evacuee and the Authority as a function of the state of nature and the actions taken. 
\begin{table*}
\caption{ Payoffs for the two players based on the state of nature and action of the Evacuee $\evac$}\label{tab:payoffs}
\[
\begin{array}{cc}
  \multicolumn{2}{c}{\evac\text{'s Payoff}} \\ 

  \multicolumn{2}{c}{\begin{array}{cccc}
&  & \multicolumn{2}{c}{\text{State of Nature}} \\ 
&  & $t$ & $n$ \\ \hline
\multirow{2}{*}{\text{$\evac$'s Action}} & L & -p\cdot d^\evac-w^\evac & -w^\evac \\ 
& S & -d^\evac & 0 \\ \hline
\end{array}
}\\
\\
 \princ\text{'s Payoff if }s=O & \princ\text{'s Payoff if }s=N \\
 
\begin{array}{ccc}
& \multicolumn{2}{c}{\text{State of Nature}} \\ 
& $t$ & $n$ \\ \hline
L & -p\cdot d^\princ-w^\princ-c & -w^\princ-c \\ 
S & -d^\princ-c & -c \\ \hline
\end{array}
& 
\begin{array}{ccc}
& \multicolumn{2}{c}{\text{State of Nature}} \\ 
& $t$ & $n$ \\ \hline
S & - d^\princ & 0 \\ 
\hline
\end{array}
\end{array}\]
\end{table*}

\begin{itemize}
  \item	$d^\evac$ is the cost to $\evac$ in terms of his/her own possible death, injury and property loss/damage. 
  \item	$d^\princ$ is the cost to $\princ$ in terms of $\evac$'s possible death, injury and property loss/damage.
  \item	$w^\evac$ is the cost of the evacuation of $\evac$ (i.e. $\evac$ leaving) to $\evac$ himself/herself. This can represent different costs such as a loss of productivity or income (typical issue of any evacuation drill/false alarm \cite{Gwynne16}), and possible legal and/or insurance-related costs (in case of a dismissed service).
  \item	 $w^\princ$ is the cost to $\princ$ of the evacuation of $\evac$. Also in this case, this relates to loss of productivity, legal/insurance costs, etc.
  \item	$c$ is the cost that $\princ$ incurs when it gives an order to evacuate $O$ (\emph{i.e.,} an evacuation order). This can be associated with the intervention of the police, emergency services, fire brigades, etc.
  \item	$p\in(0,1)$ is the probability that $\evac$ faces death, injury or property losses in case $\evac$ leaves and the state is $t$ (despite him/her leaving). 
    \end{itemize}
    It should be noted that the terms $d^\evac$ and $d^\princ$ (as well as $w^\evac$ and $w^\princ$) are not necessarily the same since they can account for different values given to life, injury or property in relation to different factors. For example, the value given to a loss of life or property by an authority might not match the value given by an evacuee due to ethical reasons, material vs affective value, \emph{etc.}

    \section{Model solution}\label{sec:solution}
We use Sequential Equilibrium (SE) \cite{Kreps} as a solution concept. In most cases though, the weaker concept of Perfect Bayesian Equilibrium (PBE) \cite{Fudenberg} can be used instead. In particular, there is one specific case in which the solution concept Sequential Equilibrium is needed to identify plausible posterior beliefs $\alpha$ for the Evacuee $\evac$. This is the case in which the Authority $\princ$ never orders an evacuation ($m_n=m_t=0$). We also further refine our set of equilibria by applying the intuitive criterion \cite{Cho}. The concept of sequential equilibrium has been been widely adopted in applications of game theory, and the intuitive criterion is arguably the most common refinement thereof (see \emph{e.g.} \cite{Gibbons} for an introduction). Experimental support for the relevance of sequential equilibrium is provided by \cite{Camerer,Brandts,Banks}. The latter two studies also indicate that the intuitive criterion is relevant when players have had time to learn and adapt their behavior.

\subsection{Best response of the Authority}
We first obtain the best response of the Authority based on its posterior belief $\gamma_i$ that the state of nature is $\omega=t$ conditional on having observed $x_i$ ($i\in\{n,t\}$). The expected payoff to the Authority choosing $s=O$ after having observed the signal $x_i$ is presented in equation \eqref{payoffauthO}.

\begin{equation}\label{payoffauthO}
u_\princ (O|x_i)=\gamma_i [r(-p d^\princ-w^\princ )+(1-r)(-d^\princ )]+(1-\gamma_i )[r(-w^\princ )+(1-r)  \cdot 0]- c
\end{equation}

The expected payoff of choosing $s=N$ after having observed the signal $x_i$ is presented in equation \eqref{payoffauthN}.

\begin{equation}\label{payoffauthN}
u_\princ (N|x_i) = \gamma_i (-d^\princ )+(1-\gamma_i )\cdot 0
\end{equation}

The payoff of choosing $s=O$ is strictly larger than the payoff of choosing $s=N$ if and only if:
\begin{equation}
\gamma_i (1-p)rd^\princ>rw^\princ+c
\end{equation}

Thus the best response of the Authority as a function of its posterior belief $\gamma_i$, conditional on having observed $x_i$ ($i\in\{n,t\}$), is

\begin{eqnarray}\label{bestresponseA}
  m_i=1	&\text{only IF} & \gamma_i\geq \frac{rw^\princ+c}{r(1-p)d^\princ }\nonumber\\
  m_i\in(0,1)	&\text{only IF} & \gamma_i=\frac{rw^\princ+c}{r(1-p)d^\princ} \\
  m_i=0	& \text{only IF} & \gamma_i\leq\frac{rw^\princ+c}{r(1-p)d^\princ}.\nonumber
\end{eqnarray}

A first observation concerning the behaviour of the Authority is the following fact. 
\begin{observation}\label{observation}
  When $\tau>1/2$, we have that $\gamma_t>\gamma_n$. So, equation \eqref{bestresponseA} implies that when $\tau>1/2$ it must be that $m_t\geq m_n$, in a best response of the Authority. 
  In particular, if the Authority's strategy is to Order an evacuation with positive probability after observing the low-threat signal $x_n$ ($m_n>0$), then it must be that it certainly Orders an evacuation under the high-threat signal $x_t$ ($m_t=1$). Similarly, if the Authority's strategy is to not Order an evacuation with certainty under the high-threat signal ($m_t<1$), then it must be that it certainly does Not order an evacuation under the low-threat signal ($m_n=0$).

  Therefore, we can summarise the Authority's strategy $(m_n,m_t)$ by using only one variable: the total probability to order an evacuation $M\in[0,1]$. Equation \eqref{totalprobbr} shows how to calculate the Authority's strategy $(m_n,m_t)$ as a function of $M$.

\begin{equation}\label{totalprobbr}
m_{t}=\left\{ 
\begin{array}{ll}
\frac{M}{y_{t}} & \text{if }M\leq y_{t} \\ 
1 & \text{if }M>y_{t}%
\end{array}%
\right. \quad m_{n}=\left\{ 
\begin{array}{ll}
0 & \text{if }M\leq y_{t} \\ 
\frac{M-y_{t}}{y_{n}} & \text{if }M>y_{t}%
\end{array}%
\right. 
\end{equation}
\end{observation}

\subsection{Best response of the Evacuee}
The strategy of the Evacuee is simply a probability $r\in[0,1]$ to Leave after receiving the evacuation order. The Evacuee's best response is based on his/her belief $\alpha$ that the state is $\omega=t$. This best response is given in equation \eqref{bestresponseE}.

\begin{eqnarray}\label{bestresponseE}
  r=1 &	\text{only IF} &	\alpha\geq \frac{w^\evac}{(1-p)d^\evac}\nonumber\\
  r\in(0,1) &\text{only IF} & \alpha=\frac{w^\evac}{(1-p)d^\evac}\\
  r=0& \text{only IF} & \alpha \leq \frac{w^\evac}{(1-p)d^\evac}\nonumber
\end{eqnarray}

As long as $M>0$, the Evacuee's posterior belief can be calculated using Bayes' rule, see equation \eqref{posterior}.

\begin{equation}\label{posterior}
  \alpha=\frac{\pi(\tau m_t+(1-\tau)m_n )}{M}
\end{equation}

Given Observation \ref{observation} above, it is possible to express $\alpha$ as a function of $M\in(0,y_t]$ (see equation \eqref{alphaasM}). 

\begin{equation}\label{alphaasM}
\alpha=\left\{%
\begin{array}{ll}
\gamma_t & \text{if }M\leq y_t \\ 
\gamma_n +\frac{\pi(1-\pi)(2\tau-1)}{y_n}\frac{1}{M} & \text{if } M>y_t%
\end{array}%
\right.
\end{equation}

\subsection{Results}\label{sec:results}
This section presents the results of the evacuation game. As a first step all possible sequential equilibria (including equilibria that arise under non-generic parameter configurations) are presented. Subsequently, a process of equlibrium refinement has been conducted using the intuitive criterion \cite{Cho}, as well as by ruling out non-generic parameter configurations. 
Formally, we rule out a strategy profile as being non-generic if it is a part of sequential equilibrium only under a set of parameter vectors (configurations) that are of zero Lebesgue measure in the parameter space. 
Intuitively, generic parameter configurations are those that do not require ``delicate'' conditions to hold among the various parameters (for example, there is no reason why $\frac{\wtildee}{1-p}$ should be exactly equal to $\gamma_t$).
Similarly, we rule out sets of beliefs of Lebesgue measure zero, when beliefs are not derived from equilibrium strategies via Bayesian updating. 
\subsubsection{All equilibria}\label{sec:allequilibria}
In this section we present all possible sequential equilibria, including all equilibria under non-generic parameter configurations, and equilibria that might fail the intuitive criterion. We explain how these are refined in the following subsection. To simplify the notation, we define the following variables $\wtildep=w^\princ/d^\princ$, $\wtildee=w^\evac/d^\evac$,  and $\ctilde=c/d^\princ$.
\begin{enumerate}
  \item\label{case1} $M=1$ ($\alpha =\pi $), $r=1$: In this equilibrium the Authority orders the evacuation regardless of the observed signal, so that the Evacuee's posterior and prior belief are the same. The Evacuee Leaves. This equilibrium exists only if 
\begin{equation*}
\pi \geq \frac{\tilde{w}^{\evac}}{1-p}
\end{equation*}%
and 
\begin{equation*}
\gamma _{n}>\frac{\tilde{w}^{\princ}+\tilde{c}}{1-p}.
\end{equation*}

\item\label{case2} $M=1$ ($\alpha =\pi $), $r\in (0,1)$: In this equilibrium the Authority orders the evacuaation regardless of the ovserved signal. The Evacuee's prior and posterior beliefs are the same. The Evacuee leaves with some positive probability $r\in(0,1)$. This equilibrium exists only if 
\begin{equation*}
r=p
\end{equation*}%
\begin{equation*}
\pi =\frac{\tilde{w}^{\evac}}{1-p}
\end{equation*}%
and 
\begin{equation*}
\gamma _{n}>\frac{\tilde{w}^{\princ}+\tilde{c}}{\tilde{w}^{\evac}}%
\pi =\frac{\tilde{w}^{\princ}+\tilde{c}}{1-p}.
\end{equation*}%
As $\gamma _{n}<\pi $, for this strategy profile to be an equilibrium, it must be that 
\begin{equation*}
\tilde{w}^{\evac}>\tilde{w}^{\princ}+\tilde{c}\Rightarrow \tilde{w}^{\evac}>\tilde{w}^{\princ}
\end{equation*}%
which means that the Evacuee values their work relatively more than the Authority does.

\item\label{case3} $M\in \lbrack y_{t},1)$ ($\alpha =\gamma _{n}+\frac{\pi (1-\pi )(2\tau
-1)}{y_{n}}\frac{1}{M}$), $r=1$: In this equilibrium the Authority always orders an evacuation after observing $x_t$ and with some  probability $m_n\in[0,1)$ after observing $x_n$. The Evacuee updates his/her belief according to Bayes's rule. The Evacuee Leaves. This equilibrium exists only if 
\begin{equation*}
M=p
\end{equation*}%
\begin{equation*}
\gamma _{n}=\frac{\tilde{w}^{\princ}+\tilde{c}}{1-p}
\end{equation*}%
\begin{equation*}
p\geq y_{t}
\end{equation*}%
and 
\begin{equation*}
p\left( \frac{\tilde{w}^{\evac}}{\tilde{w}^{\princ}+\tilde{c}}%
-1\right) \leq \frac{(1-\pi )(2\tau -1)}{1-\tau }.
\end{equation*}

\item\label{case4} $M\in \lbrack y_{t},1)$ ($\alpha =\gamma _{n}+\frac{\pi (1-\pi )(2\tau
-1)}{y_{n}}\frac{1}{M}$), $r\in (0,1)$: In this equilibrium the Authority always orders an evacuation after observing $x_t$ and with some probability $m_n\in[0,1)$ after observing $x_n$. The Evacuee updates his/her belief according to Bayes's rule. The Evacuee Leaves with some positive probability $r\in(0,1)$. This is the case of the cry wolf effect. This equilibrium exists only if 
\begin{equation*}
r=\frac{\tilde{c}}{(1-p)\gamma _{n}-\tilde{w}^{\princ}}
\end{equation*}%
\begin{equation*}
\alpha =\frac{\tilde{w}^{\evac}}{1-p}
\end{equation*}%
\begin{equation*}
M=\frac{\pi (1-\pi )(2\tau -1)}{\frac{\tilde{w}^{\evac}}{1-p}y_{n}-\pi
(1-\tau )}
\end{equation*}%
and 
\begin{equation*}
\frac{\tilde{w}^{\princ}+\tilde{c}}{1-p}<\gamma _{n}<\frac{\tilde{w}^{%
\evac}}{1-p}<\gamma _{t}.
\end{equation*}%

\item\label{case5} $M=y_{t}$ ($\alpha =\gamma _{t}$), $r=1$: In this equilibrium the Authority orders an evacuation if and only if it observes $x_t$. The Evacuee's posterior belief is the same as the Authority's. The Evacuee Leaves. This is the case of full compliance. This equilibrium exists only
if 
\begin{equation*}
\frac{\tilde{w}^{\princ}+\tilde{c}}{\gamma _{n}}>1-p\geq \max \left\{ 
\frac{\tilde{w}^{\evac}}{\gamma _{t}},\frac{\tilde{w}^{\princ}+%
\tilde{c}}{\gamma _{t}}\right\}.
\end{equation*}

\item\label{case6} $M=y_{t}$ ($\alpha =\gamma _{t}$), $r\in (0,1)$: In this equilibrium the Authority orders an evacuation if and only if it observes $x_t$. The Evacuee's posterior belief is the same as the Authority's. The Evacuee Leaves with some positive probability $r\in(0,1)$. This equilibrium
exists only if 
\begin{equation*}
  r\in\left[\frac{\ctilde}{\wtildee-\wtildep},1\right)
\end{equation*}%
\begin{equation*}
p<y_{t}
\end{equation*}%
\begin{equation*}
  \alpha=\gamma_t=\frac{\wtildee}{1-p}
\end{equation*}%
and 
\begin{equation*}
  \ctilde+\wtildep<\wtildee\leq 1-p.
\end{equation*}

\item\label{case7} $M\in (0,y_{t})$ ($\alpha =\gamma _{t}$), $r=1$: In this equilibrium the Authority orders an evacuation with some positive probability $m_t\in(0,1)$ after observing $x_t$ and does not order an evacuation after observing $x_n$. The Evacuee's posterior belief is the same as the Authority's. The Evacuee Leaves. This equilibrium
exists only if 
\begin{equation*}
M=p
\end{equation*}%
\begin{equation*}
p<y_{t}
\end{equation*}%
\begin{equation*}
\gamma _{t}=\frac{\tilde{w}^{\princ}+\tilde{c}}{1-p}
\end{equation*}%
and 
\begin{equation*}
\tilde{w}^{\princ}+\tilde{c}\geq \tilde{w}^{\evac}.
\end{equation*}

\item\label{case8} $M\in (0,y_{t})$ ($\alpha =\gamma _{t}$), $r\in (0,1)$: In this equilibrium the Authority orders an evacuation with some positive probability $m_t\in(0,1)$ after observing $x_t$ and does not order an evacuation after observing $x_n$. The Evacuee's posterior belief is the same as the Authority's. The Evacuee Leaves with some positive probability $r\in(0,1)$. This
equilibrium exists only if 
\begin{equation*}
r=\frac{\tilde{c}}{\tilde{w}^{\evac}-\tilde{w}^{\princ}}
\end{equation*}%
\begin{equation*}
M=\frac{\tilde{w}^{\evac}-\tilde{w}^{\princ}}{\tilde{c}}p
\end{equation*}%
\begin{equation*}
\gamma _{t}=\frac{\tilde{w}^{\evac}}{1-p}
\end{equation*}%
\begin{equation*}
\tilde{w}^{\princ}+\tilde{c}<\tilde{w}^{\evac}
\end{equation*}%
and 
\begin{equation*}
p<\frac{\tilde{c}}{\tilde{w}^{\evac}-\tilde{w}^{\princ}}y_{t}.
\end{equation*}

\item\label{case9} $M=0$ ($\alpha $ determined in Sequential Equilibrium), $r=1$:
In this equilibrium the Authority does not order an evacuation regardless of the observed signal. The Evacuee holds some (off-the-equilibrium-path) belief $\alpha$ determined by use of Sequential equilibrium. Sequential Equilibrium dictates that plausible beliefs $\alpha$ are only those $\alpha \in \lbrack \gamma _{n},\gamma _{t}]$. The Evacuee would Leave if he/she were to receive an evacuation order. Notice, though, that such an order is never given in equilibrium. 
This equilibrium exists only if 
\begin{equation*}
\alpha \in \left[ \max \left\{ \frac{\tilde{w}^{\evac}}{1-p},\gamma
_{n}\right\} ,\gamma _{t}\right] 
\end{equation*}%
and
\begin{equation*}
\frac{\tilde{w}^{\evac}}{1-p}\leq \gamma _{t}\leq \frac{\tilde{w}^{%
\princ}+\tilde{c}}{1-p}.
\end{equation*}

\item\label{case10} $M=0$ ($\alpha $ determined in Sequential Equilibrium), $r\in (0,1)$:
In this equilibrium the Authority does not order an evacuation regardless of the observed signal. The Evacuee holds some (off-the-equilibrium-path) belief that after receiving an evacuation order, the state is $t$ with probability $\alpha=\wtildee/(1-p)$. The Evacuee would Leave with some positive probability $r\in(0,1)$ if he/she were to receive an evacuation order. Notice, though, that such an order is never given in equilibrium. 
This equilibrium exists only if 
\begin{equation*}
\alpha =\frac{\tilde{w}^{\evac}}{1-p}
\end{equation*}%
\begin{equation*}
r\in \left( 0,\min \left\{ \frac{\tilde{c}}{\gamma _{t}(1-p)-\tilde{w}^{%
\princ}},1\right\} \right] 
\end{equation*}%
\begin{equation*}
\tilde{w}^{\princ}\leq \gamma _{t}(1-p)\leq \tilde{w}^{\princ}+\frac{%
\tilde{c}}{r}
\end{equation*}%
and 
\begin{equation*}
\gamma _{n}(1-p)\leq \tilde{w}^{\evac}\leq \gamma _{t}(1-p).
\end{equation*}

\item\label{case11} $M=0$ ($\alpha $ determined in Sequential Equilibrium), $r=0$:
In this equilibrium the Authority does not order an evacuation regardless of the observed signal. The Evacuee holds some (off-the-equilibrium-path) belief $\alpha$ determined by use of Sequential equilibrium. Sequential Equilibrium dictates that plausible beliefs $\alpha$ are only those $\alpha \in \lbrack \gamma _{n},\gamma _{t}]$. The Evacuee would Stay if he/she were to receive an evacuation order. Notice, though, that such an order is never given in equilibrium. 
This equilibrium exists only if 

\begin{equation*}
\alpha \in \left[ \gamma _{n},\min \left\{ \gamma _{t},\frac{\tilde{w}^{%
\evac}}{1-p}\right\} \right] 
\end{equation*}%
and 
\begin{equation*}
\gamma _{n}\leq \frac{\tilde{w}^{\evac}}{1-p}.
\end{equation*}
\end{enumerate}

\subsubsection{Refinement of equilibria}\label{sec:refinement}
We first refine the equilibria with $M=0$ by using the intuitive criterion \cite{Cho} to pin down the players' beliefs. This criterion is applied to our case in order to rule out equilibria which are such that if the Authority deviates and gives an evacuation order then the Evacuee assigns positive probability on the authority having observed $x_n$, despite the fact that an authority who has observed $x_n$ would not find it beneficial to give an evacuation order even if the evacuee would decide to Leave for sure after receiving $O$, and an authority who has observed $x_t$ would find it beneficial to give an evacuation order if the Evacuee was to Leave for sure after receiving $O$.

In particular, when $\gamma_n\leq \frac{\wtildep}{1-p}\leq \gamma_t$, the only plausible beliefs (according to the intuitive criterion) are $\alpha=\gamma_t$, since only the $x_t$-type Authority has a possible incentive to order an evacuation. In this case, the only remaining equilibria with $M=0$  are:

\begin{itemize}
  \item[9'.]\label{case9prime} $M=0$, $r=1$, $\alpha=\gamma_t$ (when $\gamma_t=\frac{\tilde w^%
\princ+\tilde c}{1-p}>\max\left\{\gamma_n,\frac{\tilde w^\mathbf{E}}{1-p}%
\right\}$) 

\item[10'.]\label{case10prime} $M=0$, $r\in[0,\frac{\tilde c}{(1-p)\gamma_t-\tilde w^\princ}%
]$, $\alpha=\gamma_t$ (when $\gamma_t=\frac{\tilde w^\mathbf{E}}{1-p}>\frac{%
\tilde w^\princ+\tilde c}{1-p}>\gamma_n$) 

\item[11'.]\label{case11prime} $M=0$, $r=0$, $\alpha=\gamma_t$ (when $\frac{\tilde w^\mathbf{E}%
}{1-p}>\gamma_t>\frac{\tilde w^\princ+\tilde c}{1-p}>\gamma_n$)
\end{itemize}

The cases 9' and 10' are non-generic among the above cases. Therefore, this means that if $\gamma_n<\frac{\wtildep+\ctilde}{1-p}<\gamma_t$, the only equilibrium under generic parameter configurations where both Authority types would not give an evacuation order is 11', \emph{i.e.,} the case where $\frac{\wtildee}{1-p}>\gamma_t$. In this case the Evacuee can never be convinced to leave.

In a second step, we consider all equilibria and we proceed with further refinement by identifying and ruling out non-generic configurations. Cases 2, 3, 6, 7, and 8 are non-generic in the sense that they require certain specific conditions among the different parameters which are unlikely to hold. We also argue that case 10 is non-generic as it involves the Evacuee holding knife-edge beliefs about the state of nature off the equilibrium path. Case 1 involves the Authority giving an evacuation order all the time and the Evacuee being willing to leave based on the prior alone. This is quite implausible because this means that the threat is so big and the signal technology so inaccurate that both parties would rather have the Evacuee Leave in the first place.

\begin{table}[!ptb]
  \caption{Equilibria under different parameter configurations. The top row and the leftmost column list parameter ranges while the remaining cells describe the possible equilibria (strategies $M$ and $r$ of the Authority and the Evacuee, respectively, as well as the Evacuee's posterior belief $\alpha$) for the corresponding combinations of parameters. Case numbers refer to the equilibria of sections \mbox{\ref{sec:solution}.\ref{sec:results}.\ref{sec:allequilibria}--\ref{sec:refinement}}.}\label{tab:equilibria}
\begin{equation*}
\begin{tabular}{c||cc|c|cc}
& \multicolumn{2}{c|}{$\frac{\tilde w^\princ+\tilde c}{1-p}>\gamma_t$} & 
$\frac{\tilde w^\princ+\tilde c}{1-p}\in(\gamma_n,\gamma_t)$ & 
\multicolumn{2}{c}{$\frac{\tilde w^\princ+\tilde c}{1-p}<\gamma_n$} \\ 
\hline\hline
\multirow{3}{*}{ $\frac{\wtildee}{1-p}>\gamma_t$ } & \multicolumn{2}{c|}{$M=0
$} & $M=0$ & \multicolumn{2}{c}{$M=0$} \\ 
& \multicolumn{2}{c|}{$\alpha\in[\gamma_n,\gamma_t]$} & $\alpha=\gamma_t$ & 
\multicolumn{2}{c}{$\alpha\in[\gamma_n,\gamma_t]$} \\ 
& \multicolumn{2}{c|}{$r=0$} & $r=0$ & \multicolumn{2}{c}{$r=0$} \\ 
& \multicolumn{2}{c|}{(Case 11)} & (Case 11') & \multicolumn{2}{c}{(Case 11)} \\ 
\hline
\multirow{3}{*}{$\frac{\wtildee}{1-p}\in(\pi,\gamma_t)$} & 
\multicolumn{2}{c|}{$M=0$} & $M=y_t$ & \multicolumn{2}{c}{$M=M^*>y_t$} \\ 
&  &  &  & \multicolumn{2}{c}{$\alpha=\frac{\tilde w^\mathbf{E}}{1-p}$} \\ 
& $\alpha\in\left(\frac{\tilde w^\mathbf{E}}{1-p},\gamma_t\right)$ & $%
\alpha\in\left(\gamma_n,\frac{\tilde w^\mathbf{E}}{1-p}\right)$ &  & 
\multicolumn{2}{c}{$r=\frac{\tilde c}{(1-p)\gamma_n-\tilde w^\princ}$}\\
&  &  &  & \multicolumn{2}{c}{(Case 4)}
\\ \cline{5-6}
\multirow{3}{*}{$\frac{\wtildee}{1-p}\in(\gamma_n,\pi)$} &  &  & $%
\alpha=\gamma_t$ & $M=M^*>y_t$ & $M=1$ \\ 
& $r=1$ & $r=0$     &  & $\alpha=\frac{\tilde w^\mathbf{E}}{1-p}$ & $\alpha=\pi$\\
&  &  &  & $r=\frac{\tilde c}{(1-p)\gamma_n-\tilde w^\princ}$ & $r=1$ \\ 
& (Case 9) & (Case 11) &  & (Case 4) & (Case 1)\\ 
\cline{1-3}\cline{5-6}
\multirow{3}{*}{$\frac{\wtildee}{1-p}<\gamma_n$} & \multicolumn{2}{c|}{$M=0$}
& $r=1$ & \multicolumn{2}{c}{$M=1$} \\ 
& \multicolumn{2}{c|}{$\alpha\in[\gamma_n,\gamma_t]$} &  & 
\multicolumn{2}{c}{$\alpha=\pi$} \\ 
& \multicolumn{2}{c|}{$r=1$} &  & \multicolumn{2}{c}{$r=1$} \\ 
& \multicolumn{2}{c|}{(Case 9)} & (Case 5) & \multicolumn{2}{c}{(Case 1)} \\ 
\end{tabular}
\end{equation*}
\end{table}

Table \ref{tab:equilibria} presents all sequential equilibria that survive the intuitive criterion under the various generic parameter configurations. The value of $M^*$ that appears in the table is given in Equation \eqref{emstar}. 

\begin{equation}\label{emstar}
  M^*=\frac{\pi(1-\pi)(2\tau-1)}{\frac{\wtildee}{1-p} y_n-\pi(1-\tau)}
\end{equation}
Note that $M^*\in(y_t,1)$ for the parameter configurations for which it is relevant. In these cases, the Authority is very prone to order an evacuation.

The cry wolf effect is present in situations where realizations of both signals $x_n$ and $x_t$ induce the authority to give an evacuation order with some positive probability ($M>y_t$) and the Evacuee does not always play $L$ (Leave) when receiving the order of evacuation $O$ \emph{i.e.} case~4. In particular, this happens in cases where (i) $\frac{\wtildep+ \ctilde}{1-p}<\gamma_n$ and (ii) $\frac{\wtildep}{1-p}\in(\gamma_n,\gamma_t)$. The first condition indicates that the authority wants the Evacuee Leave regardless of whether it obtained signal $x_n$ or signal $x_t$. The second condition indicates that the Evacuee is willing to leave if he/she believes that the state is $t$ with probability $\gamma_t$, and is willing to stay if he/she believed that the state is $t$ with probability $\gamma_n$. This leads to inefficiencies as the interests of the Authority and the Evacuee are not perfectly aligned.

In contrast, the ideal situation is the one where the Authority gives the order to evacuate if, and only if he/she receives the threat signal $x_t$  ($M=y_t$), and the Evacuee always Leaves when he/she receives the evacuation order ($r=1$). This is the case in which $\frac{\wtildep+ \ctilde}{1-p}\in(\gamma_n,\gamma_t)$ and $\frac{\wtildee}{1-p}<\gamma_t$, \emph{i.e.} case~\ref{case5}.

\subsection{Illustrative case study}

In order to illustrate our model's results, we provide an example parameterisation
of the model. Table \ref{tab:param} shows the values of all parameters of the model for the case study. It should be noted that the parametric values of the case study are selected to reflect a possible scenario, but they have not be linked to a specific type of evacuation \emph{(e.g.} a fire or a terrorist threat) in order to avoid a mis-use of the assumptions adopted. Our analysis takes three different values of the signal accuracy parameter $\tau$ as each value leads to a different prediction according to the model. These values can reflect different types of issues such as the type of technology employed to detect the evacuation \emph{(e.g.} a smoke alarm or a camera) and its reliability.

\begin{table}[!ptb]
  \caption{Parameter values for the case study}\label{tab:param}%
  \centering%
  \vspace{-1em}
\[  \begin{array}{cccccccc}
    \multicolumn{2}{c}{\text{Monetary variables}} &\qquad &\multicolumn{2}{c}{\text{Probability variables}} &\qquad & \multicolumn{2}{c}{\text{Accuracy cases}}\\
    \text{Variable} & \text{Value (in \$)} & &\text{Variable} & \text{Value} &  &\text{Variable} & \text{Value}\\
    \cline{1-2}\cline{4-5}\cline{7-8}
    d^\evac     & 1.5\times 10^6	& & 	\pi & 5\times10^{-4} & &\tau_1 & .7\\
 d^\princ 	& 1.5\times 10^6	& & 	p   & 1\times10^{-2} & &\tau_2 & .8\\
 w^\evac 	& 2.0\times 10^2	& &	    &         & &\tau_3 & .9\\
 w^\princ 	& 1.0\times10^2		& &     \\
 c 		&  2.0\times 10^1			\\
\end{array}
\]
\end{table}

The parameterisation uses $d^\princ=d^\evac$ with values being in the ballpark of the estimated value of statistical life \cite{Viscusi,Hultkrantz} \emph{i.e.,} the chosen values reflect credible values adopted in practice. Values for $w^\evac$ and $w^\princ$ are taken to represent the daily income of a worker (who corresponds to the evacuee) and the average daily profit per employee of a firm \emph{(i.e.} per evacuee), respectively. In this case, the cost $c$ represents the cost of ordering an evacuation \emph{per employee.} The parameters for $\pi$ and $p$ reflect that the state is, indeed, a threat with a probability of 0.1\% and that there is still a 1\% chance that an employee may suffer damages despite him/her attempting to evacuate. Such value is purely hypothetical and it has been chosen to reflect a scenario in which the chances to have negative consequences in case of evacuation are low. We are therefore assuming that the evacuation occurs in an effective manner with a low probability to reach untenable conditions.

Our analysis explores the model's predictions under different values of the accuracy parameter $\tau$. 
When the accuracy is low, $\tau=\tau_1=0.7$, then $(1-p)\gamma_n>\wtildee>\wtildep+\ctilde$ and we are in case \ref{case1} in which the authority always orders an evacuation --- even after observing $x_n$ --- the evacuee always evacuates. In this case, the threat detection technology available is not adequate.
When the accuracy is high, $\tau=\tau_3=0.9$, then $\wtildee>\wtildep+\ctilde>(1-p)\gamma_n$ and we are in case \ref{case5} in which the authority orders an evacuation if and only if signal $x_t$ is observed, the evacuee always complies with the order. Here, the threat detection technology is sufficiently good to ensure full compliance. 
Finally, when the accuracy takes the intermediate value, $\tau=\tau_2=0.8$, then $\wtildee>(1-p)\gamma_n>\wtildep+\ctilde$, we are in case \ref{case4}, in which the cry wolf effect is observed: the authority orders an evacuation whenever $x_t$ is observed but also some times when $x_n$ is observed. The evacuee responds by complying only partially. The threat detection technology cannot induce full compliance as the authority cannot determine with high enough confidence whether there is an actual threat or not.

\section{Discussion}
This section analyses all generic cases and discusses their implications. The ``bad,'' generic cases where the Authority never gives an evacuation order are the following. The first one involves the cost of giving the evacuation order being too high (case 9). This is quite implausible since if that was the case, there should not be an option to give an evacuation order to begin with. The other one (case 11) is one where the Evacuee holds implausible beliefs ($\gamma_n\leq\alpha\leq \frac{\wtildee}{1-p}\leq\gamma_t$; implausible because the Authority could induce different beliefs and be better off).

It should be noted that in most cases where $r<1$, we have that $\wtildee>\wtildep$ (the only exception arises in case 11 where $O$ is never used by the Authority in equilibrium). This means that in these cases the Evacuee values his work relatively more than the Authority does. This relates to cases where the Evacuee is highly committed in the investment made for his work (i.e. the case in which the activity and the productivity associated with it has a high cost to the Evacuee when interrupted). A typical example for such scenario is the Costa Concordia evacuation \cite{Kvamme} in which the Authority assessment of the situation lead to not order the evacuation in time to avoid casualties. Even under such conditions, if $p$ is low enough, or $\tau$ is high enough, the ``ideal'' situation 5 can be achieved. The model presented also allows us to reflect on how different types of authorities perform different evaluations of $\wtildep$ \emph{(e.g.,} a public body such as a police force might have a different evaluation than a private owner). 

In addition, the cost associated with an evacuation order for the Authority and an Evacuee clearly changes in relation to the nature of the threat scenario. A terrorist threat in a transient space (e.g. a transportation terminal such as an airport or a train/metro station) would most likely be associated with $\wtildee<\wtildep$, i.e., a relatively lower cost for the Evacuee's Leave decision for the Evacuee (as the loss of productivity is lower) than the Authority (who instead would have to take political/legal responsibility for such decision). Therefore, no cry wolf effect should be observed in such cases.

Based on the previous considerations, cases 4 and 5 are among the most interesting ones. Case 4 involves the Authority giving an evacuation order in some instances even if the signal received is $x_n$. This is the case of the cry wolf effect: the Authority's technology, threat assessment capabilities and information available is relatively good but not good enough to persuade the Evacuee that the threat is high enough for him/her to leave. In addition, the Evacuee values their work considerably (relatively) more than the Authority. Therefore the Authority, in order to make the Evacuee leave more often, gives an evacuation order $O$ even in some of the cases when he/she observes $x_n$. It should be noticed that for this to be the case, we need that the Authority actually wants the Evacuee to leave independent of his/her information (i.e. the main scope of the Authority is instruction compliance). The ``ideal'' situation is case number 5 where the Authority gives an evacuation order if, and only if he/she receives the signal $x_t$ and the Evacuee evacuates for sure when he/she receives the evacuation order.

The proposed model focuses on the study of the interaction between the Authority's decision and the subsequent Evacuee's response to either leave or stay. Therefore, the model does not consider the case of an authority ``pushing'' an evacuation to get 100\% compliance once an Evacuee has decided to stay, as this would result in case which does not consider the interactions between the authority's evacuation orders and Evacuee's reactions to them. Similarly, the model does not take into consideration the case in which an Evacuee takes a decision to evacuate on its own (i.e. without Authority's action) as the focus of the paper is to isolate the relationship between Authority's order and subsequent Evacuee's actions. 

Despite these limitations, the model presented in this paper represents an important step to evaluate how egress drills can be associated with the cry wolf effect. The question which arises is what changes can be made to remedy the cry wolf effect, i.e., to turn a cry wolf situation into an ``ideal'' one in which order compliance is high. Looking at the model results, all that is needed is to decrease $\gamma_n$, so that it is below $\frac{\wtildep+ \ctilde}{1-p}$. In this way, an Authority receiving a low threat signal loses his/her incentive to order an evacuation. In response to that, the Evacuee always evacuates when receiving the evacuation order, knowing that it must have been given by the $x_t$ Authority type. For $\gamma_n$ to be reduced, what needs to be done is to increase the precision of the signal, i.e. $\tau$. Therefore, if the Authority can obtain a clearer signal and the Evacuee knows that, then the cry wolf effect can be avoided. Interestingly, increasing the cost $c$ of giving an evacuation order can also lead to the cry wolf effect being avoided. This corresponds to an increase in the credibility of the evacuation order given by the Authority but it is wasteful, in contrast to increasing the accuracy of threat detection. The key implication of the model is that the Authority should always mention the danger associated with the situation (i.e. announced evacuation drills may be recommended). In this sense, our simple model suggests that unannounced drills would reduce the credibility of evacuation orders, thus their use should be carefully evaluated and they must ideally be associated with benefits other than just training and procedure assessment (for instance unannounced drills can be used for data collection on human behaviour that can be used for design and modelling purposes \cite{Kobes}). In other words, training benefits should be evaluated in light of the cry wolf efficiencies. In addition, the model suggests that the best way to avoid the (potentially large) inefficiency associated with the cry wolf effect is to invest in better detection of threat situations.

It is important to note that the proposed model includes a generic parametric analysis and an explanatory simple case study, and its applicability and validation to complex scenarios should be further evaluated in the context of existing evacuation research.
The few existing data sets on this issue generally present behavioural intentions (\emph{i.e.} hypothetical actions that people would do in case of an evacuation scenario) or post-disaster surveys for different types of evacuation scenarios \cite{Kyne,Vaiciulyte,Dash2007,Dash2000,Regnier,Kailiponi}. Unfortunately, those data sets often refer to the analysis of a scenario in isolation rather than evacuation behaviour during the passage of time and only a few studies investigate the issue of repeated evacuations (including the cry wolf effect) \cite{Dow,LeClerc}. Therefore, the proposed model represents a useful tool to look deeper into the interactions between the decisions of the evacuee and the authority's order/recommendation. Previous studies \cite{Kyne} refer to the fact that authorities who make evacuation orders are ``faced with tension between making evacuation orders based on incomplete predictions of information provided by other agencies and avoiding making a false alarm or false evacuation order.'' The model proposed in this paper provides the possibility to consider different cases, which cover different interactions between the strategies of the authority and the evacuees and the information available to them. The model also allows for investigation of the relationship between the information quality (\emph{i.e.} the accuracy of threat detection) available, previous evacuation experience and the resulting behaviour. The proposed model further allows for the study of causal links between the information quality (based on the available technology) and unnecessary evacuation experience, a well-known issue in the evacuation literature \cite{Kyne,Dow}.

Future work should investigate in depth the cases of sensitive facilities which involve an Evacuee with a dual role, i.e. he/she can decrease the consequences of the threat (e.g. a threat in a nuclear power plant in which the presence of the Evacuee can reduce its consequences \cite{Tanigawa}). Such complex cases would require a more refined representation of the costs associated with the evacuation for both the Authority and the Evacuee. The simple model proposed here also did not explicitly consider the impact of social influence in evacuation decision making among the evacuees \cite{Lovreglio16}. Future studies should evaluate the mutual relationships associated with the Leave decision among different Evacuees who may receive the same or different signals.

\section{Conclusion}
This paper presented a game-theoretic approach to investigate the cry wolf effect in emergency evacuation scenarios, presenting an evacuation game model and an example of its parameterisation. Possible equilibria have been obtained analysing the best responses of the Authority and the Evacuee. Model findings emphasises the need for a careful evaluation of the benefits associated with unannounced evacuation drills, which should go beyond staff and evacuee's training and assessment of evacuation procedure in order to counterbalance the possibly negative cry wolf effect associated with the decrease of  Evacuee's instruction compliance. In addition, increasing the accuracy of threat detection can prevent large inefficiencies associated with the cry wolf effect.

\section{Acknowledgements}
We would like to thank Dirk Helbing for his suggestions.
Mohlin gratefully acknowledges financial support from the Swedish Research Council (Grant 2015-01751), and the Knut and Alice Wallenberg Foundation (Wallenberg Academy Fellowship 2016-0156).

\medskip
\bibliographystyle{plain}

\end{document}